\documentclass[aps,prl,reprint,twocolumn,superscriptaddress]{revtex4-1}

\usepackage{amsmath} % include math package
\usepackage{amssymb} 
\usepackage{graphicx}% Include figure files
\usepackage{dcolumn}% Align table columns on decimal point
\usepackage{bm}% bold math
\usepackage[hidelinks]{hyperref}
\usepackage{color} % for colored text
\usepackage{siunitx}
\usepackage{comment}
\usepackage[T1]{fontenc} % for Michals name
\usepackage{soul,xcolor}

\newcommand{\fig}[1]{Fig.~\ref{fig:#1}}
\newcommand{\Fig}[1]{Fig.~\ref{fig:#1}}
\newcommand{\Figs}[1]{Figs.~\ref{fig:#1}}
\newcommand{\Supp}[1]{Supplement 1}
\newcommand{\mywidth}{1}

\makeatletter
\newcommand{\customlabel}[2]{%
   \protected@write \@auxout {}{\string \newlabel {#1}{{#2}{\thepage}{#2}{#1}{}} }%
   \phantom{\hypertarget{#1}{#2}}
}
\makeatother

\begin{document}
\setstcolor{red}

\title{Speckle engineering through singular value decomposition of the transmission matrix}

\author{Louisiane Devaud}
\email{louisiane.devaud@lkb.ens.fr}
\affiliation{Laboratoire Kastler Brossel, ENS-Universit\'{e} PSL, CNRS, Sorbonne Universit\'{e}, Coll\`{e}ge de France, 24 rue Lhomond, 75005 Paris, France}

\author{Bernhard Rauer}
\affiliation{Laboratoire Kastler Brossel, ENS-Universit\'{e} PSL, CNRS, Sorbonne Universit\'{e}, Coll\`{e}ge de France, 24 rue Lhomond, 75005 Paris, France}

\author{Jakob Melchard}
\affiliation{Institute for Theoretical Physics, Vienna University of Technology (TU Wien), A-1040 Vienna, Austria}

\author{Matthias K\"{u}hmayer}
\affiliation{Institute for Theoretical Physics, Vienna University of Technology (TU Wien), A-1040 Vienna, Austria}

\author{Stefan Rotter}
\affiliation{Institute for Theoretical Physics, Vienna University of Technology (TU Wien), A-1040 Vienna, Austria}

\author{Sylvain Gigan}
\affiliation{Laboratoire Kastler Brossel, ENS-Universit\'{e} PSL, CNRS, Sorbonne Universit\'{e}, Coll\`{e}ge de France, 24 rue Lhomond, 75005 Paris, France}

\date{\today}

\begin{abstract}
Speckle patterns are ubiquitous in optics and have multiple applications for which the control of their spatial correlations is essential.
Here, we report on a method to engineer speckle correlations behind a scattering medium through the singular value decomposition of the transmission matrix.
We not only demonstrate control over the speckle grain size and shape but also realize patterns with non-local correlations.
Moreover, we show that the reach of our method extends also along the axial dimension, allowing volumetric speckle engineering behind scattering layers.
\end{abstract}

\maketitle

Speckle formation is a universal feature of coherent wave dynamics in disordered systems.
It occurs whenever the phase front of a wave is randomly perturbed.
Particularly in optics this phenomenon gathered considerable interest, both for its fundamental aspects and its technological applications.
Although fully developed speckle patterns are inherently random they exhibit predictable properties~\cite{Goodman2007}.
Their field amplitudes follow a Rayleigh distribution while spatially they are correlated only locally.
The spatial extent of these local correlations is usually referred to as the speckle grain size.
These universal features make speckle patterns a valuable tool for imaging techniques~\cite{Lim2008,Mudry2012,Gateau2013}, optical tweezers~\cite{Douglass2012,Volpe2014} or random potentials for cold atoms~\cite{Lye2005,Billy2008}. 
For all these applications, being able to tune the speckle patterns' properties is highly desirable and would provide valuable degrees of control.

Recently, several methods have been developed to tailor the properties of speckle patterns.
They range from approaches to alter the patterns' statistics \cite{Bromberg2014,Bender2018} to others that target their spatial correlations \cite{Fischer2015,Guillon2017,DiBattista2016,DiBattista2018}.
While the former allow for nearly arbitrary intensity distributions to be realized, the latter focus mainly on single correlation features with limited flexibility.
More versatile techniques, such as in~\cite{Bender2019} concentrate only on intensity correlations and demand for an iterative procedure to arrive at the desired pattern.
Also, they act on a single axial plane and therefore cannot alter axial correlations. 
More flexible and direct methods to customize the correlations in speckle patterns are therefore highly desirable.

In this paper we strike a new path for engineering the spatial correlations of speckle patterns, utilizing a spatial light modulator (SLM) and a multiple scattering medium.
Our approach not only enables the volumetric control of the size and shape of the speckle grains after the medium but also allows to imprint non-local correlations.
To achieve this control we leverage the concept of the transmission matrix (TM) and its singular value decomposition (SVD). 
The TM is a powerful tool that encodes the relation between input and output modes of the scattering medium, allowing to focus light or transmit images at its output~\cite{Popoff2010,Popoff2010a,Rotter2017}.
Its SVD, a generalization of the eigen-decomposition to non-square matrices, can be used for selective focusing~\cite{Popoff2011a,Chaigne2014a,Jeong2018}, to identify open and closed channels~\cite{Vellekoop2008b,Kim2012,Yu2013,Akbulut2016},  or to find dispersion-free states from the spectral variation of the TM~\cite{Xiong2016,Ambichl2017a}. 
Here, we show that for spatially oversampled TMs that capture the local speckle correlations, the SVD also provides indirect control of these spatial correlations.
Consider, for example, a situation where field components of low spatial frequencies are more strongly transmitted through the medium.
In that case the first singular vector, with the highest transmission, will preferentially select these slowly-varying components, leading to an enlarged speckle grain size.
More generally, the full spectrum of singular vectors facilitates a flexible and customized control of the grain size.
We show further that purely computational Fourier filtering of the TM can be used to achieve arbitrary correlations, from asymmetric speckle grains to non-local correlations and Bessel-like speckle patterns.
Moreover, the control obtained through the SVD extends also to the axial direction, going beyond the reach of established methods.

\begin{figure}[t]
\centering
    \includegraphics[width=\mywidth\columnwidth]{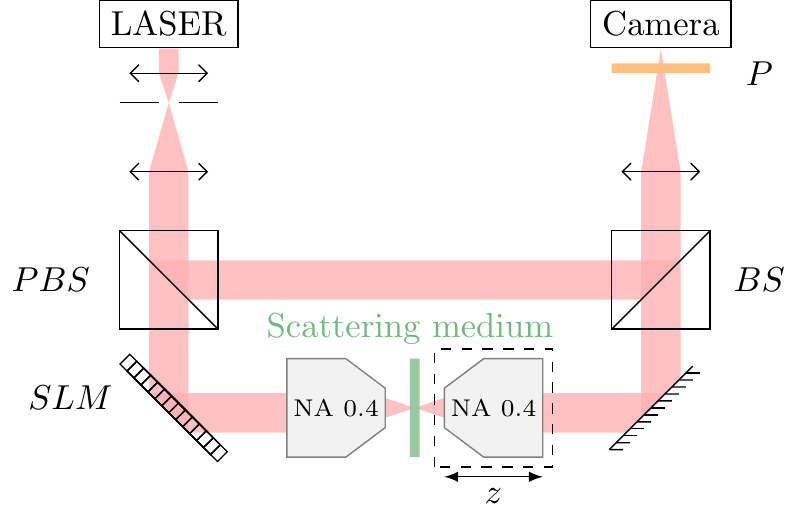}
    \caption{
    Experimental scheme. 
    A beam delivered from a Ti:sapphire laser (MaiTai HP, Spectra-Physics) is divided into two paths by a polarizing beam splitter (PBS). 
    On one path the wavefront is modulated by a reflective phase-only SLM (HSP512L-1064, Meadowlarks).
    This modulated light is focused on a layer of scattering material (TiO$_2$ layer suspended on a glass slide, thickness of $\sim \!\SI{10}{\micro \metre}$ resulting in a transmittance of $\simeq 0.3$) by an imaging objective with a numerical aperture (NA) of 0.4 (Olympus PLN20X).
    An identical second objective is used to image the transmitted scattered light onto a charged coupled device (CCD) camera (Manta G-046B, Allied Vision).
    The second objective is mounted on a translational stage allowing to image the transmitted light at varying distances $z$ from the medium.
    The two paths are recombined on a beam splitter (BS).
    A polarizer (P) in front of the camera ensures that only light with a selected polarization is measured.
    }
    \label{fig:experimental_setup}
\end{figure}

Our experimental setup is presented in \fig{experimental_setup}. 
The wavefront of a coherent light source is modulated by an SLM before being focused on a layer of scattering material. 
The scattered light that is transmitted through the medium forms a speckle pattern imaged onto a camera.
Over a second path, an unperturbed plane wave, decoupled from the beam before the SLM, is recombined and interferes with the scattered light at the CCD. 
This configuration enables the measurement of the complex light field after the medium through digital phase-stepping holography.
The polarization of the reference path is chosen to be orthogonal to the one of the light sent onto the diffusive medium, ensuring that only multiply scattered light is detected in the interferogram~\cite{xu2005random}.
The TM is constructed by displaying a basis of input patterns on the SLM and recording the corresponding output fields~\cite{Popoff2010}.

%%% Fig. 1 %%%%%%%%%%%%%%%%%%%%%%%%%%%%%%%%%%%%%%%%
%%%%%%%%%%%%%%%%%%%%%%%%%%%%%%%%%%%%%%%%%%%%%%%%%%%

\begin{figure*}[t]
\centering
    \includegraphics[width=1.\textwidth]{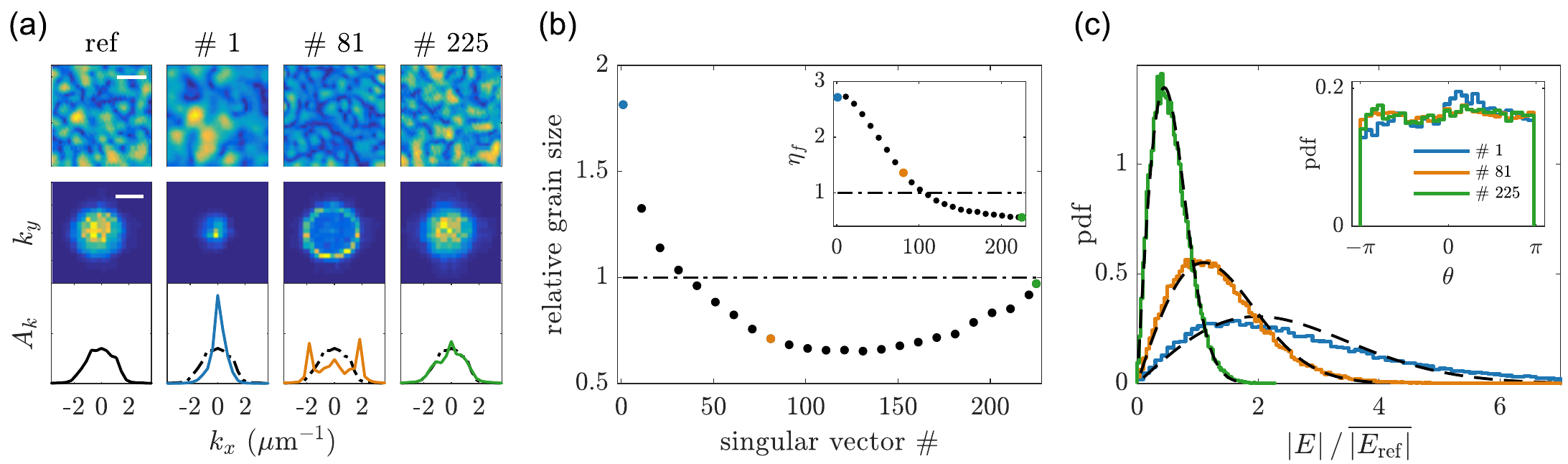}
    \caption{Speckle grain size control through the SVD of the TM. 
    (a) Speckle patterns obtained from displaying a blank reference pattern (ref) and three different singular vectors of the TM ($\#\;1$, $\#\;81$, $\#\;225$, out of 225). 
    The first row shows examples of the observed field amplitude patterns in real space while the second row displays the respective Fourier space distributions (scale bars: \SI{5}{\micro \meter} and \SI{2}{\per \micro \meter} respectively).
    The third row shows central cuts through these Fourier space distributions (blue, orange, green lines) compared to the reference (black).
    The cuts are averaged over 3 pixel rows.
    For all panels the data are individually normalized and do not reflect the changes in total transmission.
    (b) Relative speckle grain size as a function of the singular vector number (reference grain size: \SI{2}{\micro \meter}).
    The inset plots the enhancement of the field amplitude with respect to the blank reference, $\eta_f$, for the different singular vectors.
    The dotted-dashed black lines represent the mean reference values.
    The data points corresponding to the examples shown in (a) are marked with their respective colors (blue, orange and green).
    (c) Probability density functions (pdf) of the field amplitudes obtained for the three singular vectors displayed in (a). 
    All distributions are normalized to the average field amplitude of the reference $\overline{|E_\mathrm{ref}|}$ and fitted with Rayleigh distributions (dashed black lines).
    The inset shows the corresponding phase distributions.
    All results, except for speckle patterns in (a) are averaged over 36 realizations of the disorder.
    }
    \label{fig:svd_evol}
\end{figure*}

We first measure the TM at a distance of \SI{1}{\milli\meter} behind the medium output surface (see~\cite{SM},~Sec.~I). 
In order to capture the local correlations of the light field in the TM measurement we spatially over-sample the imaged speckle patterns (i.e., a single speckle grain covers several camera pixels).
Performing the SVD of the measured TM gives access to the sorted singular values and associated singular vectors.
We send one of these vectors through the medium by applying its phase as a modulation on the SLM.
To study how the speckle grain size depends on the singular vector chosen, we record the light field at the image plane.
Figure~\ref{fig:svd_evol}(a) shows the observed speckle patterns for three different singular vectors together with a reference speckle obtained when displaying a blank pattern on the SLM.
The first singular vector ($\#\;1$), associated to the largest transmission, leads to an enlarged speckle grain size, corresponding to a narrower $k$ distribution in Fourier space compared to the reference.
An intermediate vector ($\#\;81$), leads to a grain size smaller than the reference while the last vector ($\#\;225$) results in a speckle that is indistinguishable from a random input state. 
In Fourier space, increasing the vector number leads to the appearance of a ring with growing radius and decreasing amplitude (for the last vectors the ring fades completely).
Similar effects of ring-shaped distributions of light in real space have been observed for the SVD of an acousto-optic TM~\cite{Katz2019}.

%%% Fig. 2 %%%%%%%%%%%%%%%%%%%%%%%%%%%%%%%%%%%%%%%%
%%%%%%%%%%%%%%%%%%%%%%%%%%%%%%%%%%%%%%%%%%%%%%%%%%%

To quantitatively compare the speckle grain size for different inputs we calculate the output pattern's autocorrelation and extract its full width at half maximum (FWHM).
Figure~\ref{fig:svd_evol}(b) shows the speckle grain size over the singular vector number, relative to the size of the reference speckles.
The first singular vectors lead to a grain size increase whereas the intermediate ones result in a decrease, as observed in \fig{svd_evol}(a).
Selecting the singular vector number within this range therefore allows to choose patterns with a desired grain size. 
The last singular vectors display a continuous transition back to the reference grain size.
The change in grain size is accompanied by a change in overall transmission (\fig{svd_evol}(b), inset), as expected for the singular vectors \cite{Kim2012}.

The above experimental observations can be understood from the speckle Fourier space distribution obtained for the blank reference input (\fig{svd_evol}(a)).
There, we see that the distribution is peaked around $|\vec{k}|=0$ meaning that spatial modes with small transverse frequency components are transmitted better than high-frequency ones.
As the SVD returns the eigenmodes of the TM associated to specific values of transmission, it also results in a specificity in spatial frequency.
The highly transmitting first singular vectors concentrate light in the low-frequency spatial modes resulting in larger grains.
For the less transmitting singular vectors the opposite is happening, leading to a suppression of low-frequency components, smaller speckle grains and the ring shaped distributions observed in \fig{svd_evol}(a).
A model based only on a random TM with local Gaussian correlations enables to well reproduce the experimental results (see~\cite{SM},~Sec.~II).
The only condition for frequency selectivity to appear is the non-flatness of the output Fourier distribution.

Interestingly, the control gained over the local correlations of the output patterns does not substantially influence their statistics.
Figure~\ref{fig:svd_evol}(c) shows the field statistics of the three example vectors shown in \Fig{svd_evol}(a). 
In general, the field amplitudes retain their Rayleigh distribution and the phases are uniformly distributed, as expected for fully developed speckles patterns (\Fig{svd_evol}(c), inset).
Only the first singular vector shows small deviations in its amplitude distribution.
We attribute these to the high concentration of light in the forward scattering component with $|\vec{k}|=0$, as for some realizations, a preferred phase can be observed~\cite{SM},~Sec.~III.

\begin{figure}[h!]
\centering
    \includegraphics[width= \mywidth\columnwidth]{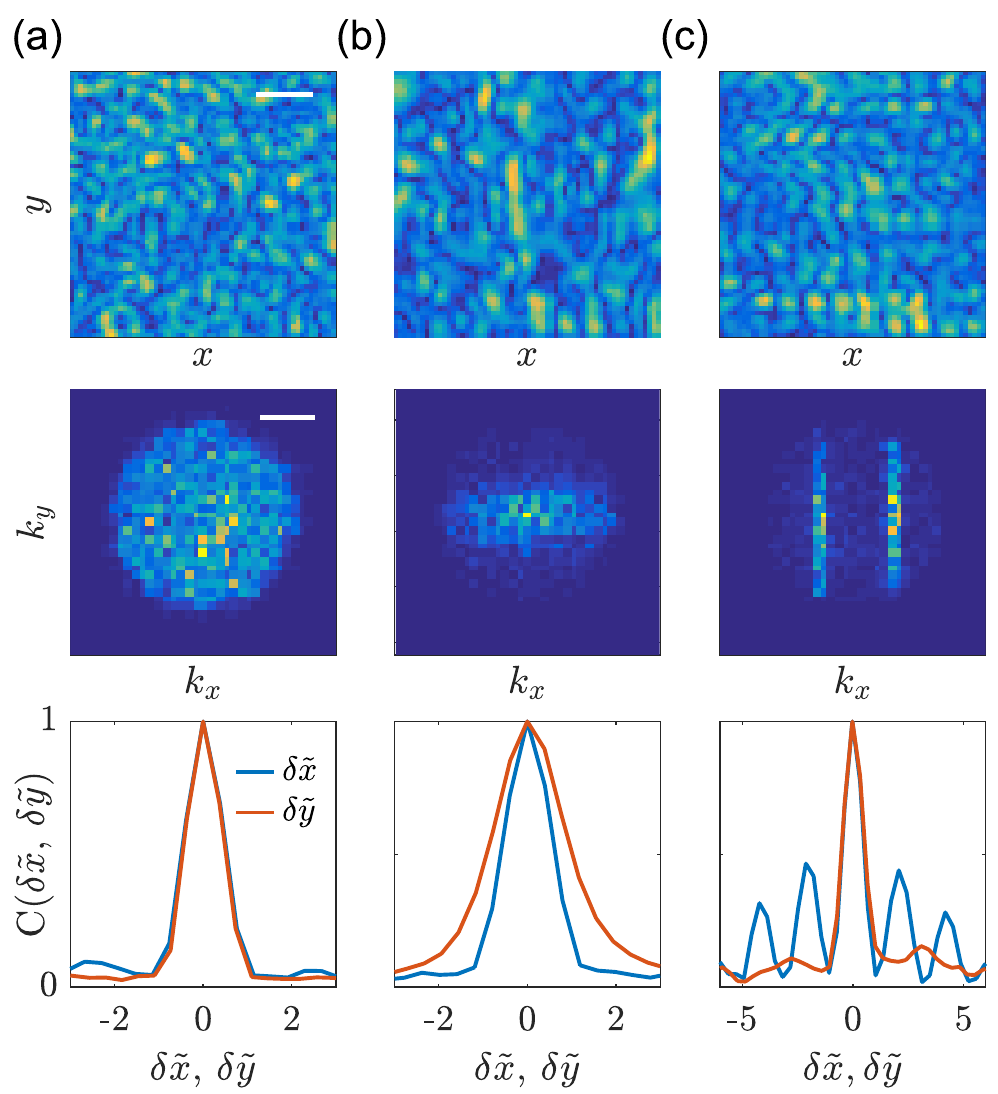}
    \caption{
    Speckle engineering through the SVD of computationally filtered TMs with initially uniform spatial distribution, imaged at the output plane of the medium.
    (a) Reference amplitude speckle (blank SLM input).
    (b) Amplitude speckle obtained from displaying the first singular vector of a TM computationally filtered by a line-shaped Fourier mask.
    (c) The same for a TM filtered by two vertical lines selecting a specific $|k_x|$-range in Fourier space.
    The first row shows a single realization of the speckle pattern, while the second row shows the distribution in Fourier space (scale bars: \SI{5}{\micro \meter} and \SI{2}{\per \micro \meter} respectively, initial grain size: \SI{1.4}{\micro \meter}).
    The third row shows the autocorrelation function $C(\delta x) = |\langle E^*(x) E(x + \delta x) \rangle/\langle  |E(x)|^2\rangle|$ along both the $x$ (blue) and $y$ (red) directions.
    The horizontal axis is rescaled by the FWHM of the reference speckle autocorrelation $w_\mathrm{ref}$, with $\delta \tilde{x} = \delta x / w_\mathrm{ref}$.
    The Fourier distributions as well as the autocorrelations represent an average over 4 disorder realizations.
    }
    \label{fig:some_masks}
\end{figure}

An important ingredient for the successful control of the grain size is the non-flatness of the speckle's spatial frequency distribution for a random input.
In our case, the limited surface area at which light exits the medium naturally leads to a peaked distribution when imaging a plane at a distance from the surface~\cite{Goodman2007}.
This becomes clear when picturing the limiting case of imaging a plane far removed from the medium. There, the exit surface would act like a point source and all light would converge to the $k = 0$ mode (see~\cite{SM},~Sec.~I).
If, instead, we image the speckles right at the medium output, the collected $k$ vectors are mainly limited by the imaging NA, leading to a uniform distribution of spatial frequencies, as shown in \fig{some_masks}(a).
In such a situation, the singular vectors show no dispersion in grain size.

A convenient way to overcome such limitations is to artificially introduce the non-flatness in the spatial frequency distribution.
This can be done by employing the approach developed in \cite{Boniface2017}, where Fourier filtering of the TM was used to shape the point spread function when focusing light behind a medium.
A TM is experimentally measured, its output dimension is Fourier transformed and multiplied with a filter mask that introduces the desired variation from the flat distribution.
After transforming it back to real space, the SVD of this filtered TM can be performed.
Note that this is a fully computational procedure and no changes to the scattering medium nor to the measurement configuration are necessary~\cite{SM},~Sec.~IV.

The large flexibility of the filtering process provides a wide range of control over the speckle correlations.
Figure~\ref{fig:some_masks}(b) and (c) show the results of two masks showcasing the potential of this approach. 
The first mask filters the high $k_y$ components without modifying the $k_x$ components (\fig{some_masks}(b)).
This results in a speckle pattern where grains are elongated along $y$ while their extension in $x$ remains unchanged (for a different method to achieve this asymmetry see~\cite{SM},~Sec.~V).
The second mask selects only a specific $|k_x|$-range leading to two symmetric stripes in Fourier space (\fig{some_masks}(c)).
Correspondingly, the speckles exhibit a short range periodicity.
The filtering approach thus enables not only to shape the speckle grains almost arbitrarily but also to imprint non-local correlations on them.

%%% Fig. 3 %%%%%%%%%%%%%%%%%%%%%%%%%%%%%%%%%%%%%%%%
%%%%%%%%%%%%%%%%%%%%%%%%%%%%%%%%%%%%%%%%%%%%%%%%%%%

A particularly interesting application is to realize speckles with a Bessel-shaped autocorrelation.
For that, spatial frequencies need to be concentrated on a ring, as for regular Bessel beams \cite{McGloin2005}.
The latter gathered considerable interest due to their non-diffractive nature and several techniques were developed to generate Bessel beams behind scattering media~\cite{DiBattista2015,DiBattista2016,Boniface2017}.
Bessel speckle patterns on the other hand remain little studied although their enhanced depth-of-field could be advantageous in structured illumination microscopy or speckled optical potentials \cite{Schwartz2007}.

To experimentally realize a Bessel speckle pattern we first measure the TM at an arbitrary position $z_0$ behind the medium and apply the filtering technique using a ring shaped mask.
This provides us with a filtered TM that only encodes the transmission of spatial frequencies of a certain absolute value $|\vec{k}|$.
As before, we perform the SVD of this filtered TM and display the phase of the first singular vector on the SLM. 
To verify the Bessel-like characteristics of the generated speckles we record their axial features by measuring the speckled light field while scanning the positions $z$ of the imaged plane around $z_0$, providing a volumetric measurement of the output field.

\begin{figure}[t]
\centering
    \includegraphics[width=\mywidth\columnwidth]{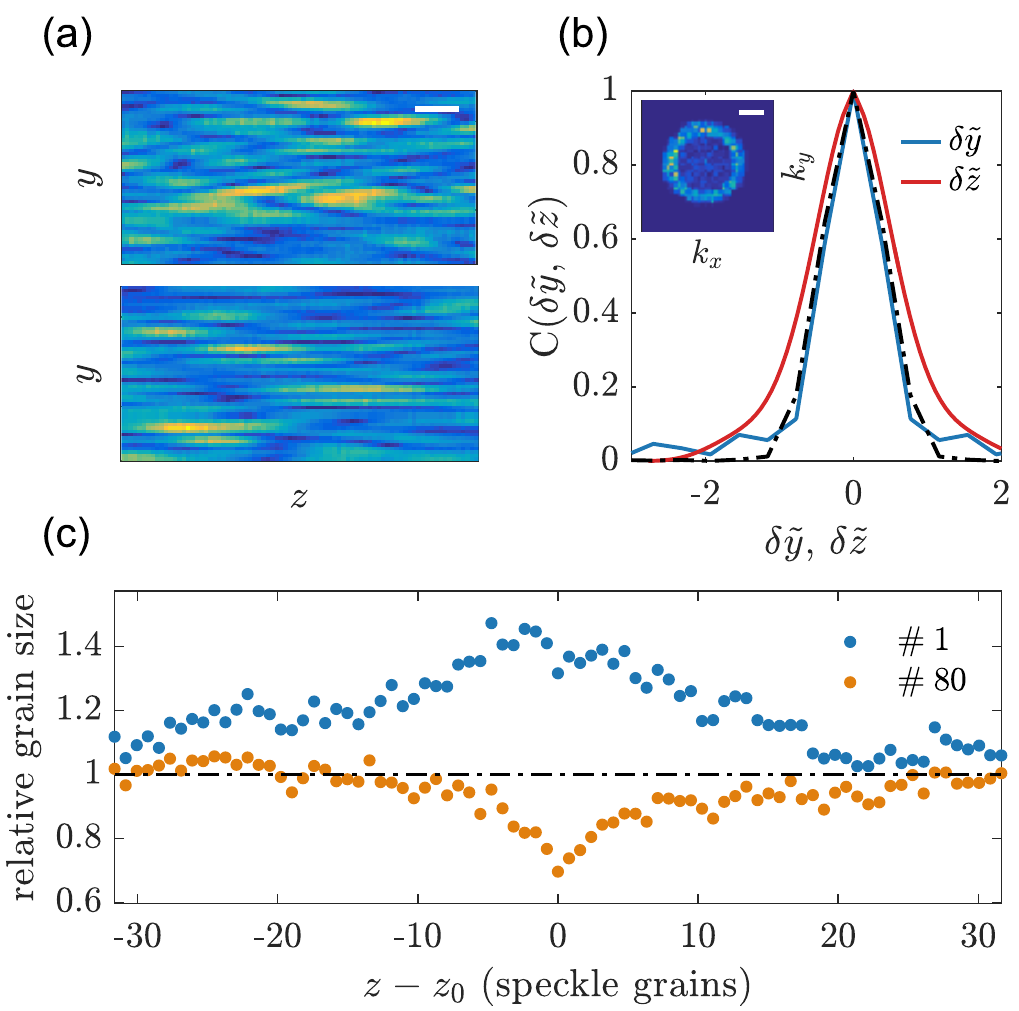}
    \caption{
    Speckle correlation control along the axial dimension.
    (a) Reference amplitude speckle pattern (top) and output obtained for the first singular vector of a ring filtered TM aimed at creating a Bessel speckle pattern (bottom).
    Both show cuts in the $(y,z)$ plane for an arbitrary $x$ position (scale bar: \SI{5}{\micro \meter}).
    (b) Plot of the autocorrelation $C$ in $y$ (blue) and $z$ (red) direction for the first singular vector of a ring-filtered TM.
    Again, for both curves, the horizontal axis is rescaled by the FWHM of the reference autocorrelation along the respective dimensions (see \fig{some_masks}).
    The dash-dotted black line indicates the reference autocorrelation (along $y$, equivalent to the one along $z$ due to the rescaling) obtained from a blank input.
    Autocorrelation data are averaged over all $x$ positions.
    The inset shows the corresponding Fourier space distributions in the $(x,y)$ plane at the position $z_0$ (scale bar: \SI{2}{\micro \meter^{-1}}).
    (c) Relative transverse speckle grain size along the axial dimension $z$ for data taken under similar conditions as in \fig{svd_evol}. 
    Shown are the first singular vector ($\#\;1$, blue) and an intermediate one ($\#\;80$, orange).
    The $z$-axis is rescaled by the axial extension of the reference speckle at $z_0$, where the TM was measured (\SI{12}{\micro \meter}).
    Note that the diffraction induced increase of the speckle grain size with $z$ is not visible here as only values relative to the local reference grain size are given.
    }
    \label{fig:bessel}
\end{figure}

The results are presented in \fig{bessel}(a) and (b), where we observe that the axial extension of the speckle grains indeed increases.
Along the transverse directions, the limited spatial resolution prevents us from observing a full Bessel-shaped autocorrelation.
Yet, a slight decrease in width can be qualitatively observed and the spatial frequencies of the output patterns indeed show the desired concentration on a ring.

%%% Fig. 4 %%%%%%%%%%%%%%%%%%%%%%%%%%%%%%%%%%%%%%%%
%%%%%%%%%%%%%%%%%%%%%%%%%%%%%%%%%%%%%%%%%%%%%%%%%%%

This result is particularly interesting as it shows that our method allows to control the speckle grain size in all three spatial dimensions.
Even at planes away from where the TM was measured, spatial correlations are affected.
Since our method does not rely on interference, changes in spatial correlations do not depend on a complicated superposition of modes that align properly only in a single plane.

Correspondingly, the ring in Fourier space selected by the first singular vector (\fig{bessel}(b), inset) remains visible over several axial grains and the speckle grain modulations presented earlier persist over a large volume.
Figure~\ref{fig:bessel}(c) shows the relative transverse grain size along the axial dimension $z$, measured under similar conditions as the data in \fig{svd_evol}.
For the first singular vector we observe the grain size enhancement over a range of about 20 axial grains.
In case of the intermediate singular vector the axial range of control is smaller but still extends beyond a few grains in $z$-direction.
These observations reach significantly beyond \cite{Bender2019} where the customization of speckle correlations is valid in a single plane only.

In conclusion, we experimentally demonstrate a novel method to engineer speckle correlations behind scattering media.
We show how the SVD of a TM with local correlations leads to a dispersion of singular modes with respect to the output speckle grain size if the initial transmission of spatial frequencies is non-uniform.
Fourier-filtering of the TM can further be used to achieve arbitrary spatial mode distributions, enabling asymmetric speckle grains, non-local correlations or Bessel-like speckle patterns.
The filtering even lifts the requirement of oversampling the TM, facilitating its practical application.
Note that the range of control obtained for the speckle correlations depends on the amount of input modes controlled at the SLM compared to the region-of-interest size at the output.
While the experiments discussed here have been realized on small regions of 100 to 400 grains, they can be scaled up by increasing the number of controlled modes~\cite{SM},~Sec.~VI.
Compared to previous approaches, our method proves to be more flexible and achieves a larger axial range of control.
Possible applications range from structured illumination microscopy to the engineering of optical random potentials or algorithmic advantages in computational imaging~\cite{Chen2014,Wang2020a}.
Moreover, on a fundamental level, the principle of SVD-based control can be translated to other platforms that employ matrices encoding input-output relations as in acoustics \cite{aubry2009random} or in integrated photonic circuits \cite{mingaleev2003scattering}.

\vspace{5mm}

\begin{acknowledgments}
We would like to thank Micha\l{} D\k{a}browski, Lorenzo Valzania, Antoine Boniface, Julien Guilbert, Jonathan Dong and Gianni Jacucci for the carefull reading of the manuscript.
This project was funding by the European Research Council under the grant agreement No. 724473 (SMARTIES), the European Union's
Horizon 2020 research and innovation program
under the Marie Sk\l odowska-Curie grant agreement No. 888707 (DEEP3P) and the Austrian Science Fund (FWF) under Project No. P32300 (WAVELAND).
SG is a member of the Institut Universitaire de France.
\end{acknowledgments}

\bibliography{Papers-TM_SVD_clean,add_bib-TM_SVD}

%%%%%%%%%%%%%%%%%%%%%%%%%%%%%%%%%%%%%%%%%%%%
%%%%%%%%%% Supplemental Materials %%%%%%%%%%
%%%%%%%%%%%%%%%%%%%%%%%%%%%%%%%%%%%%%%%%%%%%

\setcounter{secnumdepth}{3}
\clearpage
\onecolumngrid

\renewcommand{\thefigure}{S\arabic{figure}}
\renewcommand{\theequation}{S\arabic{equation}}
\setcounter{equation}{0}
\setcounter{figure}{0}
\setcounter{section}{0}
\newcommand{\Supps}[1]{Supp.~\ref{sec:#1}}

\setstcolor{red}
\customlabel{fig:experimental_setup}{1}
\customlabel{fig:svd_evol}{2}
\customlabel{fig:some_masks}{3}
\customlabel{fig:bessel}{4}

\begin{center}
  \LARGE
  \textbf{Supplemental Material: Speckle engineering through singular value decomposition of the transmission matrix}
\end{center}

\section{Experimental details}
\label{sec:exp_details}

\subsection{Geometry of illumination and detection}
\label{sec:TM_pos}

\begin{figure}[tb]
\centering
    \includegraphics[width=0.6\columnwidth]{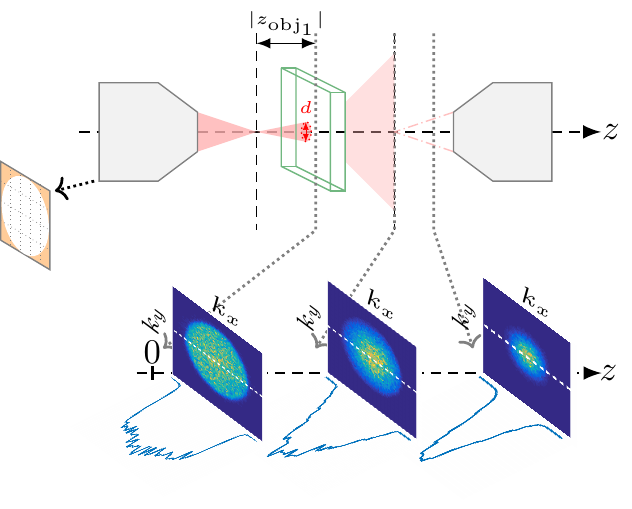}
    \caption{Top: illumination and collection geometry.
    The first microscope objective is defocused by $|z_{\mathrm{obj}_1}|$ = \SI{0.5}{\milli \meter} giving an illuminated area of $\pi d^2/4=$\SI{0.06}{\square \milli \meter}.
    The scattering medium thickness is about \SI{10}{\micro \meter}.
    The SLM is imaged on the back focal plane of the illumination microscope objective such that its circular aperture cuts off the modes at the edge of the controlled region (on the orange region).
    The imaged plane is at a varying distance $z$ from the sample.
    Bottom: spatial frequency distribution of speckles for different imaging planes behind the medium surface (positioned at $z = 0$).
    The imaged planes are, from left to right: $z \simeq 0$, $ z \simeq 0.8$ and $z \simeq$ \SI{1.2}{\milli \meter}.
    Corresponding central cuts through the distributions are shown below. 
    The shown data is obtained by averaging over 20 realizations of random input patterns displayed on the SLM.
    }
    \label{fig:supp_TM_pos}
\end{figure}

As indicated in \fig{experimental_setup} of the main text, the phase pattern on the SLM is imaged onto the back focal plane of the illumination objective such that the input surface of the scattering media is close to the Fourier plane of the SLM.
However, as presented in~\Fig{supp_TM_pos}, we slightly defocus the illumination objective by a fixed distance $|z_{\mathrm{obj}_1}|$ = \SI{0.5}{\milli \meter} to increase the illuminated area to a diameter of $d \,\simeq$ \SI{280}{ \micro \meter}.
As the medium's thickness is comparatively small ($\simeq \SI{10}{\micro \meter}$), the area at which the light exits the medium is approximately the same as the illumination area.
On the camera, its image covers the whole CCD such that different realizations can obtained simply by choosing different regions around the center of the image.
The magnification of the imaging system is 22.2.

On the output surface, the extent of the illuminated area together with the distance $z$ at which the speckle is observed determine the speckle grain size (given a blank input pattern is displayed on the SLM)~\cite{Goodman2007}.
In the experiment, the limited NA of the collection objective influences the grain size as well.
These features are best observable in Fourier space, revealing the spatial frequencies present in the patterns. 
\Fig{supp_TM_pos} (bottom part) shows the Fourier transformed speckle patterns at three imaging planes along the $z$ direction.  
Imaging the speckles right at the medium output surface, the observed spatial frequencies are only limited by the NA of the objective, leading to a flat top distribution.
Moving the imaging plane away from the medium, the limited exit surface of the light becomes dominant and the distribution gets peaked at low spatial frequencies, corresponding to an increase of the speckle grains.

For the measurements performed with an inhomogeneous speckle Fourier distribution, as in \fig{svd_evol} and \fig{bessel}(c) of the main text, we image a plane at a distance $z$ $\simeq$ \SI{1}{\milli \meter} behind the medium.
For the data presented in \fig{some_masks} and \Figs{bessel}(a-b) a plane close to the medium output surface was imaged.

%%% Fig. SM?  %%%%%%%%%%%%%%%%%%%%%%%%%%%%%%%%%%%%%%%%
%%%%%%%%%%%%%%%%%%%%%%%%%%%%%%%%%%%%%%%%%%%%%%%%%%%

\subsection{Ratio of controlled and detected modes}
\label{sec:sm_control}

The measured TMs have the dimension $N_{\mathrm{CCD}}$ $\times$ $N_{\mathrm{SLM}}$, where $N_{\mathrm{CCD}}$ is the number of pixels in the region-of-interest on the CCD and $N_{\mathrm{SLM}}$ is the number of input modes controlled by the SLM. 
For the measurements presented in \fig{svd_evol} and \fig{bessel}(c) of the main text $N_{\mathrm{SLM}} = 1024$ and $N_{\mathrm{CCD}} = 225$ while for the measurements of 
\fig{some_masks} and \Figs{bessel}(a,b) $N_{\mathrm{SLM}} = 4096$ and   or $N_{\mathrm{CCD}} =900$.
However, these numbers do not directly translate to the number of controlled and measured modes.
Due to the intentional oversampling of the measured speckles the output patterns contain fewer independent modes, specifically 100 for the smaller region-of-interest and 400 for the larger.
Furthermore, circular beam clipping at the illumination objective's back focal plane cuts the modes at the edges of the square region controlled on the SLM, see \fig{supp_TM_pos}.
Therefore, only about 80\% reach the scattering medium. 
This results in a output-to-input control ratio of 0.2 for all measurements presented.\\
To reach the previous control values some experimental precautions are taken.
The modulated SLM region covers the light beam to avoid non modulated co-propagating light.
Thus SLM pixels are grouped to form macropixels (of 10$\times$11 pixels for 1024 modes).
Also, due to the magnification, a single speckle grain covers several CCD pixels.
When measuring the TM the images are binned.
For the data of~\fig{svd_evol}, after a binning of 3$\times$3 pixels, a speckle grain corresponds to two CCD pixels to fulfill the over-sampling condition.

\section{Minimal model for the grain size control}
\label{sec:high_sv}

In order to better understand the results on the speckle grain size control reported in the main text we devise a simple model that convincingly reproduces the experimental results.
First, we generate a fully random TM with independent and identically complex Gaussian distributed elements.
Then we take each column of this TM corresponding to the output image for a given input pixel and impose local correlation.
For that we reshape the column values into a 2D image and convolve it with a Gaussian whose width defines the spatial correlation length, i.e. the speckle grain size.
Applying this convolution process to each 2D image and reshaping them back to single columns gives us the correlated TM.

\begin{figure}[h]
\centering
    \includegraphics[width=1.0\columnwidth]{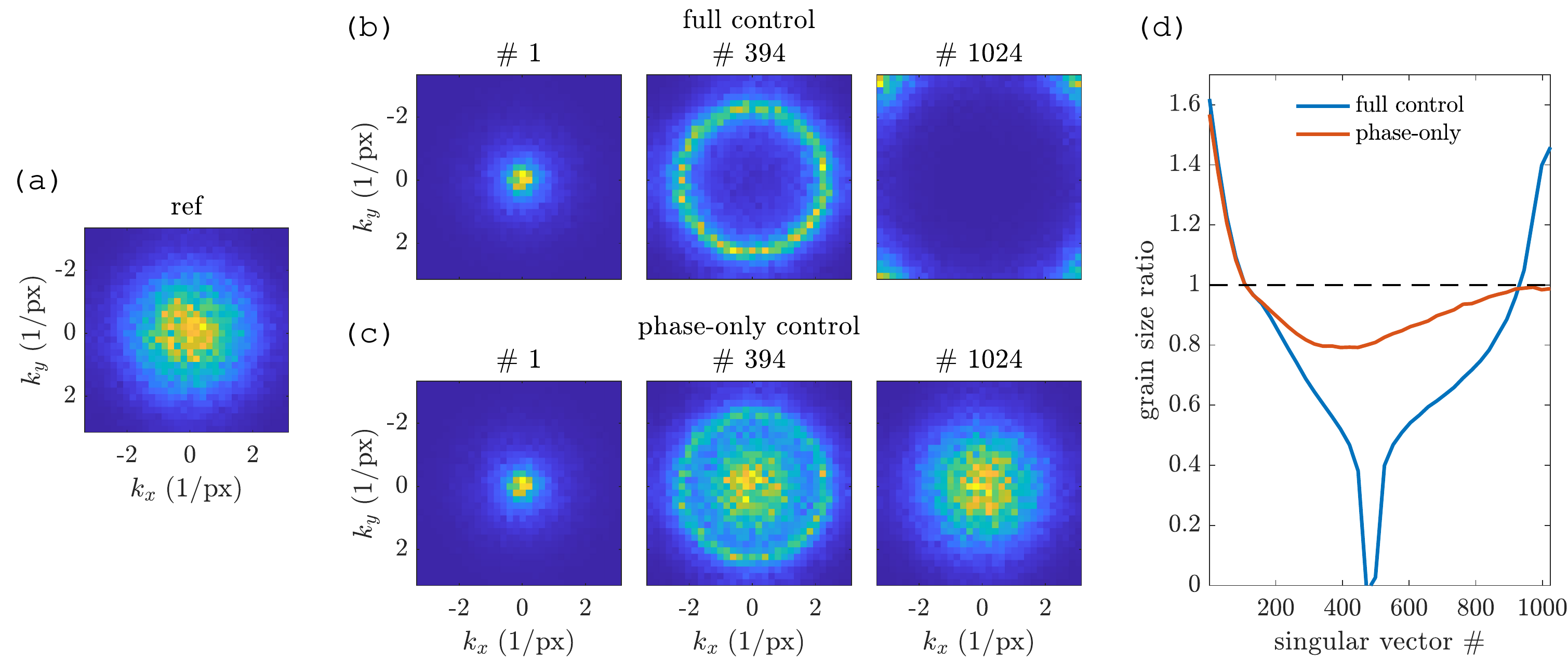}
    \caption{Results from a minimal model of a random TM with local Gaussian correlations. 
    A TM of size $1024\times1024$ was generated with a local correlation length of 2 pixels (FWHM) assuming a $32\times32$ output plane. 
    (a) Shows the reference Fourier space distribution obtained for a random input pattern.
    (b) Shows the spatial frequency distribution of the output speckles when applying the first singular vector (left), an intermediate vector in the spectrum (middle) and the last vector (right). 
    In (c) the same is shown assuming phase-only control at the input side. 
    The speckle grain size enhancement ratio, extracted as in the experimental analysis, is presented in (d) for both the cases of full control (blue) and phase-only control (red). 
    All results were averaged over 10 realizations of the TM.}
    \label{fig:supp_simulation}
\end{figure}

For this correlated TM, we can now perform the SVD and calculate the output speckle patterns corresponding to the individual singular vectors, in analogy to the experiment. 
\Fig{supp_simulation}(b) shows the distribution of spatial frequencies for the output field of three of those vectors using phase and amplitude information while \fig{supp_simulation}(c) shows the same with only the phase of the singular vectors being projected. 
In the second case, the results are in good agreement with the experimental findings shown in \fig{svd_evol}(b) of the main text.
Also, the extracted grain size presented in \fig{supp_simulation}(c) agrees well with the experimental observations in the case of phase-only control.

The Gaussian form of the local correlations used in this minimal model does not fully reflect the experimental grain size correlations.
Nevertheless, the model shows that nothing more than local correlations that result in a non-uniform distribution of spatial frequencies is needed for the effect to appear.
Important to note is that the phase-only restriction does not affect much the output for the first singular vectors.
We attribute this to the fact that the first singular vectors are transmitted better, enhancing their contribution compared to the background created by not displaying the full singular vector.

%%% Fig. SM3  %%%%%%%%%%%%%%%%%%%%%%%%%%%%%%%%%%%%%%%%
%%%%%%%%%%%%%%%%%%%%%%%%%%%%%%%%%%%%%%%%%%%%%%%%%%%

\section{Speckle statistics}
\label{sec:speckle_statistic}

\begin{figure}[tb]
\centering
    \includegraphics[width=0.5\columnwidth]{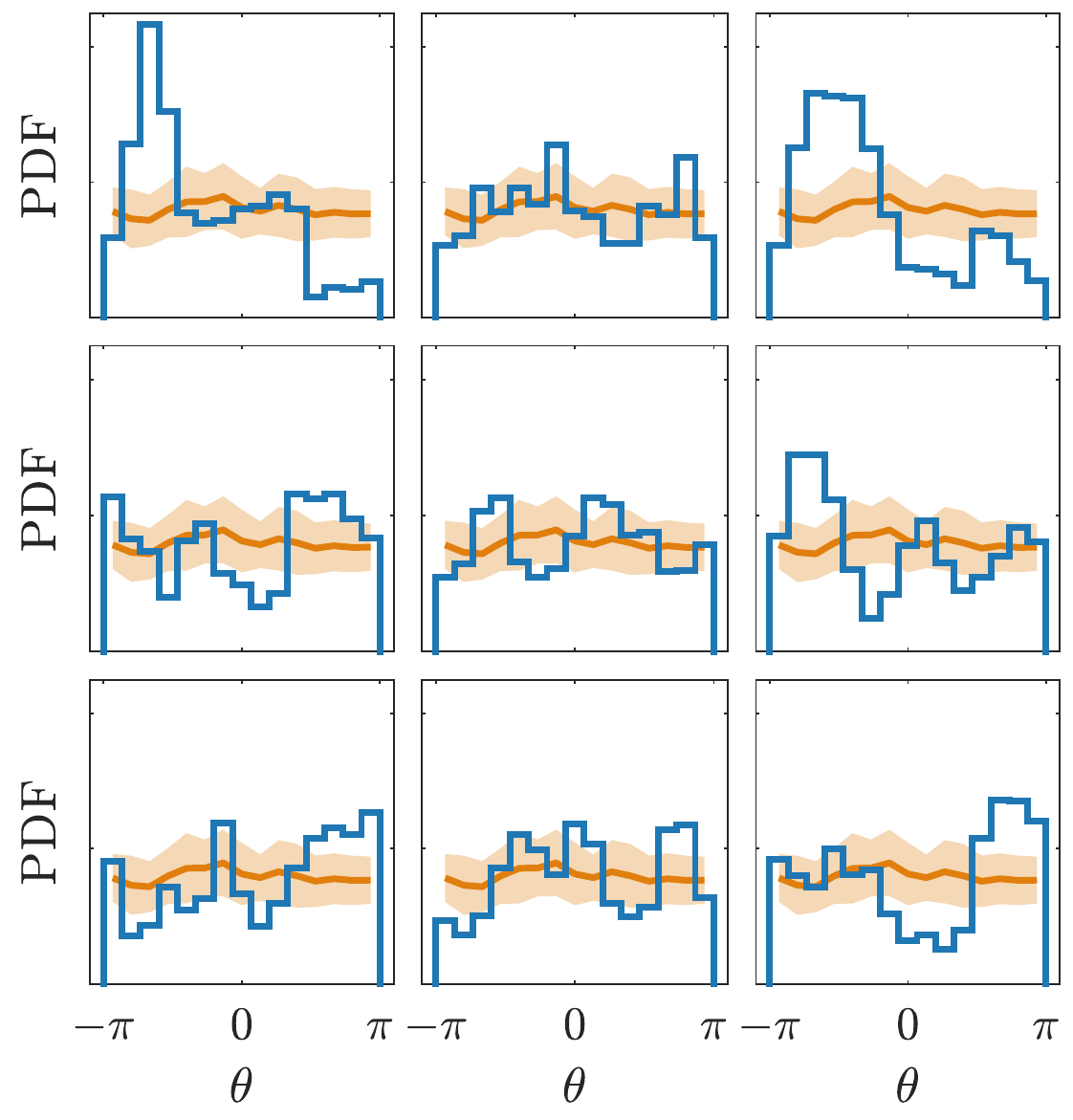}
    \caption{
    Histogram showing the phase distributions for individual realizations of the output patterns obtained from displaying the first singular vectors.
    The data is the same as shown in \fig{svd_evol}, presenting 9 of the 36 realizations.
    The orange line gives the average distribution of the reference speckle with the shaded area marking one standard deviation.
    }
    \label{fig:supp_statistic}
\end{figure}

The field amplitude statistic of a fully developed speckle pattern follows a Rayleigh distribution while its phase is uniformly distributed between $-\pi$ and $\pi$.
In the main text we pointed out small deviations from this behaviour for the output speckle amplitudes obtained from the first singular vector, shown in \fig{svd_evol}(c).
Here we perform a more thorough analysis of the phase distributions of the individual speckle realizations.
\Fig{supp_statistic} shows the individual output phase distributions of the first singular vector for a representative selection of realizations. 
For some realizations the speckles show a preferred phase, deviating from the statistics of a fully developed speckle.
As discussed in the main text, we attribute this to the enhancement of the $|\vec{k}|=0$ components of the field through the first singular vector.
Note that even though the distribution can get peaked, it still spans the full phase range, generally regarded as a hallmark of fully developed speckles \cite{Goodman2007}.
Also, since the phase peaks are randomly distributed, no preferred phase persists on average (see inset of \fig{svd_evol}(c)). 
Hence, on average also the first singular vector leads to Rayleigh distributed speckles with only small deviations of the amplitude distribution and the individual phase distributions, compatible with most applications requiring fully developed speckles.

Of the overall modes present in the medium we only control and measure a very limited fraction due to the small NA of the used objectives and the restriction to a single polarization state of the light.
Therefore we are confident that the peaked phase distributions cannot stem from open channels or mesoscopic correlations \cite{Rotter2017}.

%%% Fig. SM5  %%%%%%%%%%%%%%%%%%%%%%%%%%%%%%%%%%%%%%%%
%%%%%%%%%%%%%%%%%%%%%%%%%%%%%%%%%%%%%%%%%%%%%%%%%%%

\section{Filtering technique}
\label{sec:filtering_technique}

\begin{figure}[tb]
\centering
    \includegraphics[width=0.8\columnwidth]{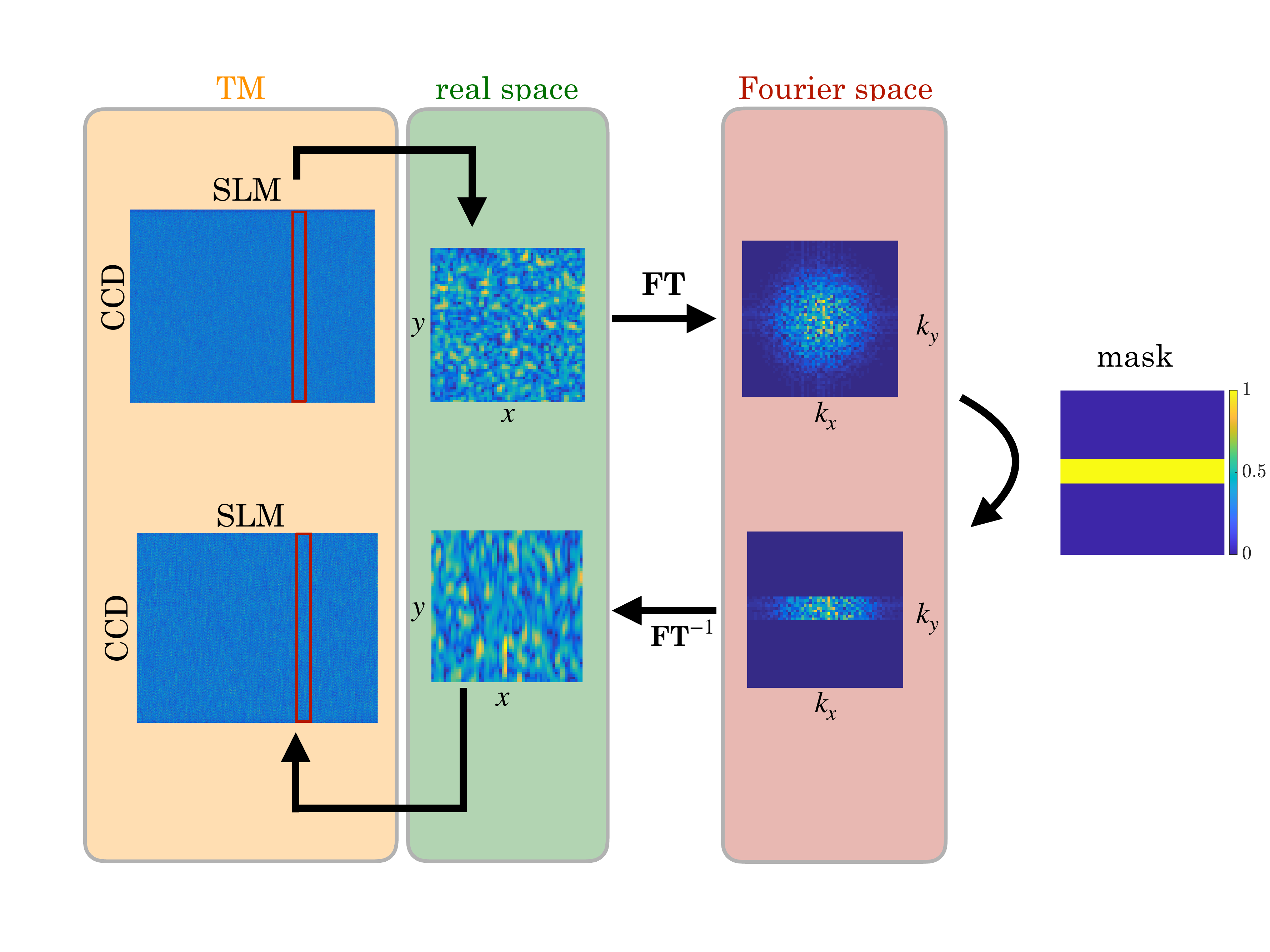}
    \caption{
    Scheme of the TM filtering technique. 
    For each input mode controlled at the SLM, the output field, represented by a column of the TM, is first reshaped in its original image configuration and then Fourier transformed (FT).
    After that, the Fourier representation is multiplied with a mask that selectively enhances or subdues certain $k$ components (in the example shown the mask is binary).
    The modified field is transformed back to real space and reshaped into a column of the now filtered TM.
    The example TM shown here represents simulated data obtained from the model discussed above.
    }
    \label{fig:filtering_technique}
\end{figure}

The filtering technique used is adapted from ref.~\cite{Boniface2017} and is presented in \fig{filtering_technique}.
This filtering is solely numeric; no experimental changes need to be made to change from one filtering mask to another.

%%% Fig. SM1  %%%%%%%%%%%%%%%%%%%%%%%%%%%%%%%%%%%%%%%%
%%%%%%%%%%%%%%%%%%%%%%%%%%%%%%%%%%%%%%%%%%%%%%%%%%%

Note that, in contrast to~\cite{Boniface2017}, no decrease in efficiency is expected when filtering away more frequency components.
In~\cite{Boniface2017}, after Fourier transforming the filtered $k$ distribution back to real space, a single pixel was selected to focus on.
However, after filtering, only a limited number of modes is available to interfere constructively at the focus site, reducing the efficiency compared to the unfiltered case.
For the speckle engineering techniques presented here, on the other hand, the first singular vector concentrates the transmitted power in the targeted spatial modes without any further restrictions. 
When filtering more, the stronger reduction in the number of output modes only results in smaller transmission increase ratios for initial singular vectors.

%%% Fig. SM1  %%%%%%%%%%%%%%%%%%%%%%%%%%%%%%%%%%%%%%%%
\begin{figure}[tb]
\centering
    \includegraphics[width=0.8\columnwidth]{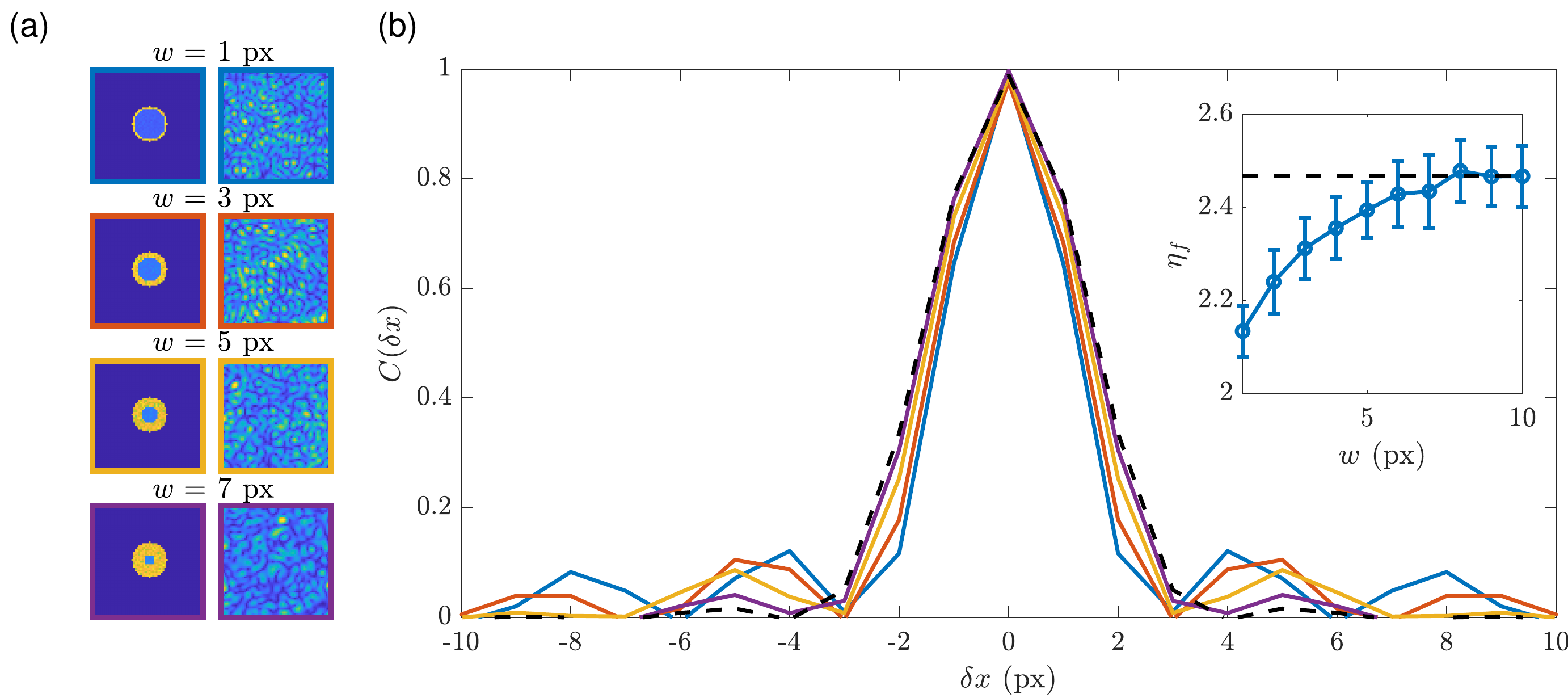}
    \caption{Simulation of the filtering step impact on the noise.
    (a) Average Fourier transform of the fields behind the scattering medium (left) for different widths $w$ of the filtering ring mask and examples of the associated speckles' field amplitudes (right).
    (b) Field autocorrelation (line colors correspond to the frame colors in (a)) and in inset the field enhancement of the first singular vector ($\#~1$) as a function of $w$ for a complete set of masks.
    Error bars are given by the field enhancement standard deviation.
    The data are averaged over 50 realisations.}
    \label{fig:filtering_technique_Bessel}
\end{figure}
%%%%%%%%%%%%%%%%%%%%%%%%%%%%%%%%%%%%%%%%%%%%%%%%%%%

To illustrate these points, we performed simulations and investigated the quality of the Bessel-like autocorrelation function for different widths $w$ of the filtered ring (see~\Fig{filtering_technique_Bessel}).
Here, no transformation back to real space after the Fourier filtering step was performed to demonstrate that in principle it is not necessary.
Stronger filtering (small values of $w$) leads to more well-defined autocorrelation features without any apparent noise or background appearing as visible in~\Figs{filtering_technique_Bessel}(a,b): the speckle grains are visually narrower and the autocorrelation presents the characteristic enhance secondary maxima of a Bessel function. 
Also the fluctuations from realization to realization are not enhanced.
However, as expected due to the reduction of the degrees of freedom in the output dimension of the filtered TM, the distribution of singular values changes and the overall transmission enhancement drops for more selective filtering masks (\Fig{filtering_technique_Bessel}(c)).

\section{Asymmetric grains through one-dimensional singular value decomposition}
\label{sec:1d_svd}

In \fig{some_masks}(b) of the main text we show how Fourier filtering can be used to create speckle patterns with asymmetric grains.
Here we propose an alternative approach to realize such patterns which does not require any Fourier filtering. 
The idea is to apply the SVD only along one spatial dimension of the output region. 
In the TM, the columns associated to the individual pixels of the output region are not necessarily ordered.
Exchanging two columns does not change the TM or the singular vector spectrum.
The information which pixel is located next to which is contained only in the correlations between the TM elements.
These correlations are also the source of the grain size dispersion observed in the output patterns of the singular vectors.
Therefore, if we remove the correlations along one spatial dimension we can achieve the grain size enhancement along the remaining one only.
To do so we divide the TM into sub-matrices which only encode the transmission for a single pixel row of the output field.
Let us consider a TM of dimension $N_{\mathrm{CCD}} \times N_{\mathrm{SLM}}$, where $N_{\mathrm{CCD}}$ is the number of pixels in the output region-of-interest and $N_{\mathrm{SLM}}$ is the number of input modes controlled by the SLM.
For simplicity we will assume a square region-of-interest such that the number of pixel rows at the output is $\sqrt{N_{\mathrm{CCD}}}$.
For each of these rows we obtain a sub-matrix of size $\sqrt{\mathrm{N_{CCD}}} \times N_{\mathrm{SLM}}$.
The SVD is now performed separately on each of these sub-matrices, removing the correlations between the pixel rows from the process.
Adding all resulting first singular vectors obtained from the individual decompositions constructs a mode that increases the grain size only along the spatial dimension associated to the output rows.
The same procedure can of course be applied to the columns as well.

\begin{figure}[tb]
\centering
    \includegraphics[width=0.4\columnwidth]{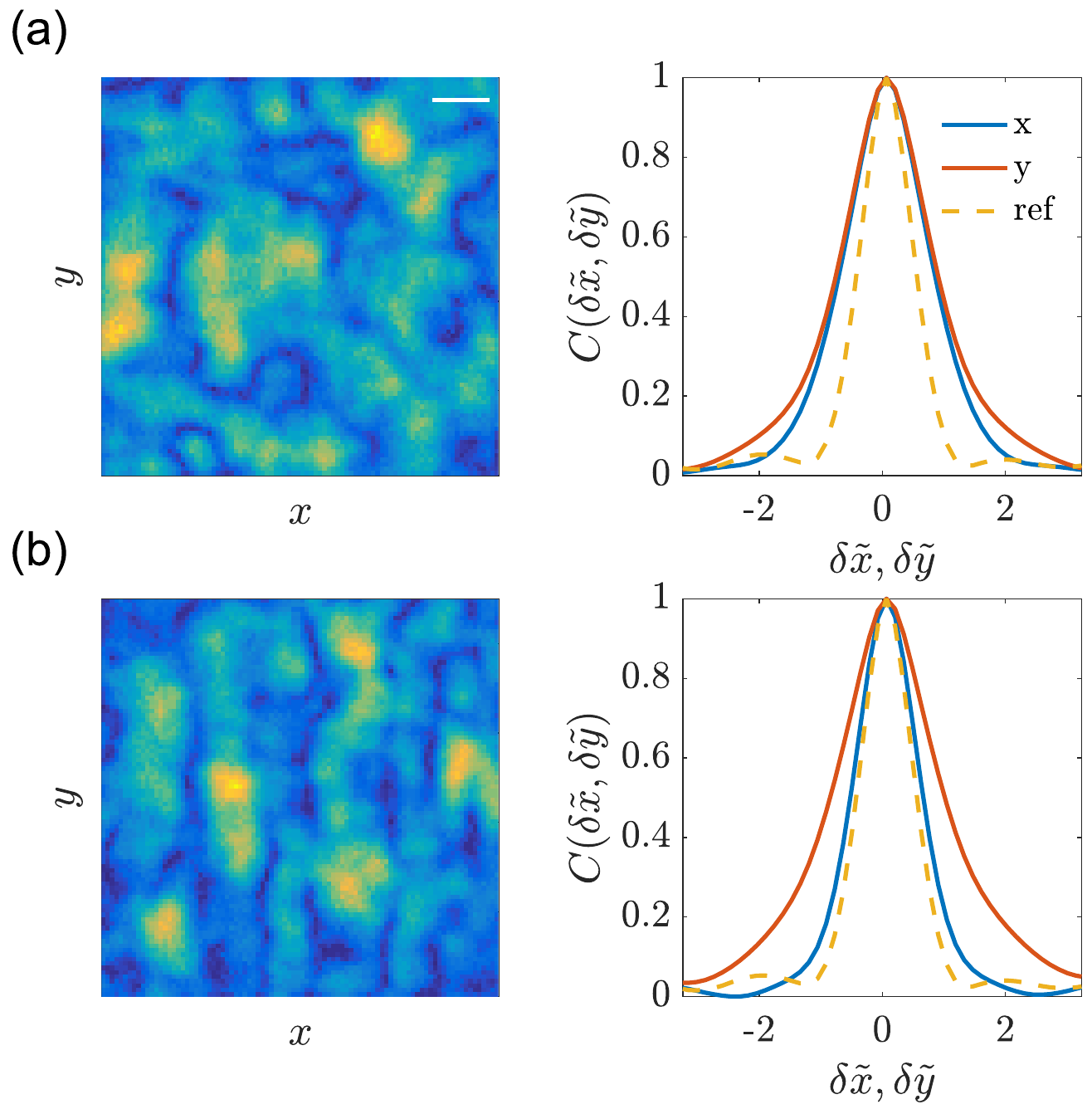}
    \caption{Elongated speckles obtained from the one-dimensional SVD.
    Panel (a) shows the output for the first singular vector obtained from the regular SVD of the TM while (b) shows the output obtained from the 1D SVD.
    For both, an example of the output speckle amplitude (left, scale bars: \SI{5}{\micro \meter}) and
    the autocorrelation (right) are given.
    The autocorrelation is plotted along the $x$ (blue) and $y$ (red) directions and is compared to the reference speckle autocorrelation (orange dashed). 
    The $x$-axes are rescaled by the FWHM of the reference speckle autocorrelation (see \fig{some_masks} caption). 
    The autocorrelation curves are averaged over 9 realizations.
    }
    \label{fig:filtering}
\end{figure}

\Fig{filtering} shows the experimentally obtained speckles from the first singular vector of the regular SVD and the 1D SVD discussed above. 
In the case of the regular SVD the grains are enlarged in both $x$ and $y$ directions while for the one-dimensional SVD the elongation only happens in one direction leading to elongated speckle grains.

%%% Fig. SM2 %%%%%%%%%%%%%%%%%%%%%%%%%%%%%%%%%%%%%%%%
%%%%%%%%%%%%%%%%%%%%%%%%%%%%%%%%%%%%%%%%%%%%%%%%%%%

This technique gives good results and is easy to implement experimentally.
The speckle elongation achieved is even better than the one obtained by the filtering technique discussed in the main text. 
However, it allows only a specific deformation of the speckle grains while the filtering technique is much more versatile and intuitive.

\section{Impact of the degree of control}
\label{sec:control_effect}

\begin{figure}[tb]
\centering
    \includegraphics[width=0.8\columnwidth]{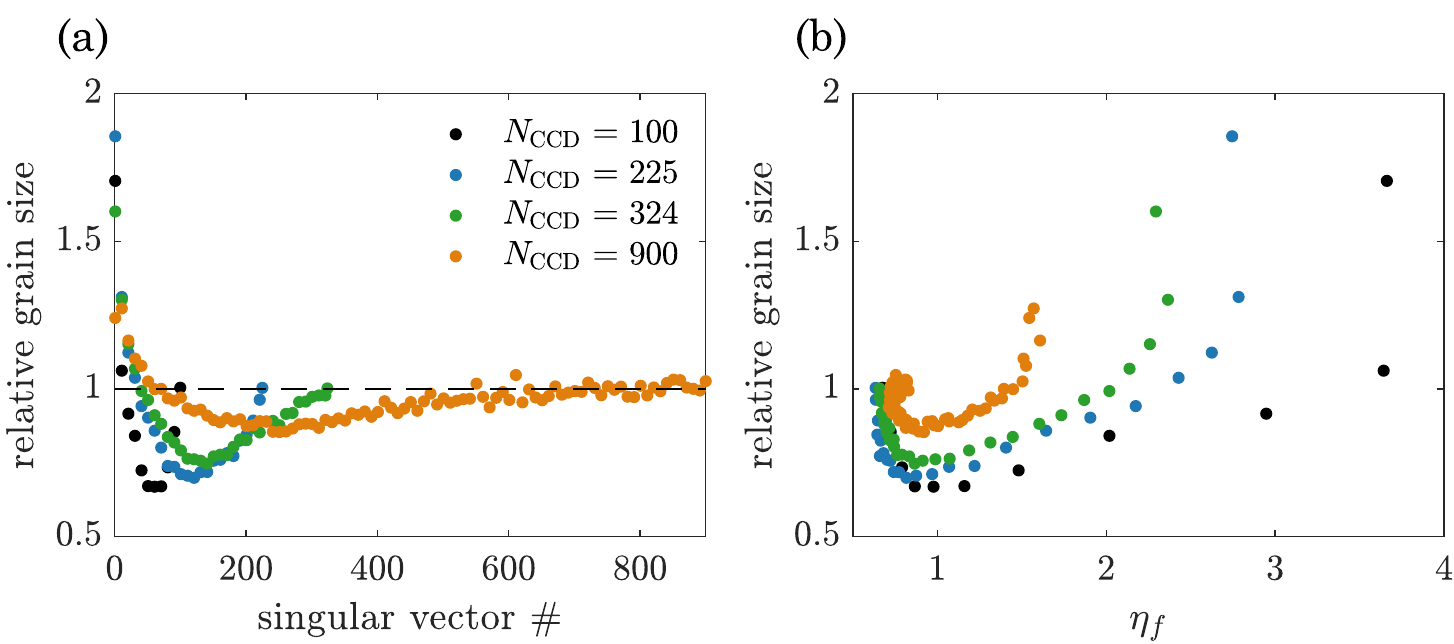}
    \caption{Impact of the degree of control on the grain size control achieved with the SVD.
    (a) Relative grain size as a function of the singular vector number with the degree of control.
    The qualitative behavior remains the same while the amount of control scales range of the grain size variation.
    (b) Relative grain size as a function of the measured field enhancement $\eta_f$.
    Note that the minimum of the curve is always obtained for $\eta_f = 1$.
    All grain sizes values are averaged over the same number of grains, i.e. for the smaller region-of-interests more realizations are averaged.}
    \label{fig:supp_control_effect}
\end{figure}

To complement the data shown in \fig{svd_evol}(b) of the main text we performed the same measurement for different amounts of control.
As in~\fig{svd_evol}, we image a plane at a distance ($z$ = \SI{1}{\milli \meter}) after the medium output surface. 
The number of modes controlled on the SLM is fixed to 1024 while a large region-of-interest is selected on the CCD camera.
This region-of-interest is then successively subdivided into smaller ones of identical sizes to perform averaging.
Measurements where taken with TMs of size $100 \times 1024$, $225 \times 1024$, $324 \times 1024$ and $900 \times 1024$.
Due to the subdivision of a fixed initial region-of-interest, the number of realizations averaged depends on the TM size (81, 36, 25 and 9 respectively), but is always realized over the same amount of speckle grains.
The obtained results of the relative grain size evolution are presented in~\fig{supp_control_effect}(a).
The overall effect we observe is a smooth decrease from enhanced grain sizes for the initial singular vectors to smaller grains for intermediate vectors, followed by a return to the reference size.
However, the amount of variation depends on the amount of control: the larger the ratio of controlled input modes to measured output modes the larger the variations achieved.
In~\fig{supp_control_effect}(b), plotting the relative grain size as a function of the field amplitude enhancement highlights an interesting point: the minimum grain size is obtained for $\eta_f =1$.
For lower enhancements $\eta_f < 1$ we observe a smooth evolution back to the reference grain size.

We observe that the exact shape of the relative grain size curves is dictated by the reference speckle $k$-space inhomogeneities.
The common feature however is the smooth variation from enhancement to reduction, followed by a return to the reference size.

Note that we defined the grain size through the width of the autocorrelation of the amplitude speckle. This is done in order to be consistent with the experimental amplitude speckles displayed. However, defining the grain size through the intensity, the field amplitude or the field itself, gives the same relative grain sizes due to the normalization by the reference grain.

%%% Fig. SM?  %%%%%%%%%%%%%%%%%%%%%%%%%%%%%%%%%%%%%%%%
%%%%%%%%%%%%%%%%%%%%%%%%%%%%%%%%%%%%%%%%%%%%%%%%%%%

\end{document}